\documentclass[pre,aps,onecolumn,superscriptaddress,nofootinbib]{revtex4-1}
\pdfoutput=1
\usepackage{amssymb}
\usepackage{epsfig}
\usepackage{amsmath}
\usepackage{subfigure}
\usepackage{graphicx}
\usepackage{textcomp}
\usepackage{url}
\usepackage{float}
\usepackage{color}
\usepackage{cases}
\usepackage[normalem]{ulem}

\usepackage[colorlinks=true, urlcolor=blue, anchorcolor=blue, citecolor=blue,filecolor=blue,linkcolor=blue,menucolor=blue%,pagecolor=blue
]{hyperref}

\usepackage{color}

\usepackage{epsfig}
\usepackage{textcomp}
\usepackage{epstopdf}

\begin{document}
\title{Large fluctuations of the area
 under a constrained Brownian excursion}

\author{Baruch Meerson}
\email{meerson@mail.huji.ac.il}
\affiliation{Racah Institute of Physics, Hebrew University of
Jerusalem, Jerusalem 91904, Israel}

\pacs{05.40.-a, 05.70.Np, 68.35.Ct}

\begin{abstract}
We study large fluctuations of the area $\mathcal{A}$ under a Brownian excursion $x(t)$ on the time interval $|t|\leq T$, constrained to stay away from a moving wall $x_0(t)$ such that $x_0(-T)=x_0(T)=0$ and $x_0(|t|<T)>0$. We focus on wall functions described by a family of generalized parabolas $x_0(t)=T^{\gamma} [1-(t/T)^{2k}]$, where $k\geq 1$.
Using the optimal fluctuation method (OFM), we calculate  the large deviation function (LDF) of the area at long times. The OFM provides a simple description of the area fluctuations in terms of optimal paths, or rays, of the Brownian motion.   We show that the LDF has a jump in the third derivative with respect to $\mathcal{A}$ at a critical value of $\mathcal{A}$.  This singularity results from a qualitative change of the optimal path, and it can be interpreted as a  third-order dynamical phase transition.

Although the OFM is not applicable for typical (small) area fluctuations, we argue that it correctly captures their power-law scaling of $\mathcal{A}$ with $T$, with an  exponent that depends continuously on $\gamma$ and on $k$. We also consider the cosine wall $x_0(t)=T^{\gamma} \cos[\pi t/(2T)]$ to illustrate a different possible behavior of the optimal path and of the scaling of typical fluctuations. For some wall functions additional phase transitions, which result from a coexistence of multiple OFM solutions,  should be possible.

\end{abstract}

\maketitle

\section{Introduction}

Brownian motion, constrained to stay away from a moving wall, is a standard setting in non-equilibrium statistical mechanics and theory of random processes. One subclass of this constrained Brownian motion is a Brownian excursion  $x\left(t\right)$, with $x \left(-T\right) \! = \! x\left(T\right) \! = \! 0$, which must stay away from a moving wall $x_0(t)$ such that $x_0(-T)=x_0(T)=0$ and $x_0(|t|<T)>0$.  Frachebourg and Martin \cite{Frachebourg2000} studied this setting in the context of the one-dimensional Burgers equation in the inviscid limit with white-noise initial condition. In that case the relevant moving wall is parabolic,
$x_{0}\left(t\right) \! = \! T^2-t^2$. The parabolic case was also studied by Groeneboom \cite{Groeneboom1989}. Ferrari and Spohn \cite{FS} considered a semicircle $x_{0}\left(t\right) \! = \! \sqrt{T^2-t^2}$, a more general parabola $x_{0}\left(t\right) \! = \! T^{\gamma}\left(1-t^{2}/T^{2}\right)$ and some other wall functions.  The authors of these works were interested in the statistical properties of typical (small) fluctuations of the Brownian particle's position away from the moving wall at a specified time $\tau \in (-T,T)$ in the limit of $T\to \infty$. The recent work \cite{SmithMeerson2018} revisited this setting in the context of atypical \emph{large deviations} of the particle away from the wall.

Here we also consider a Brownian excursion that escapes a moving wall, but suggest  a different characterization of the fluctuations away from the wall. We will be interested in the probability density $\mathcal{P}\left(\mathcal{A}, T\right)$  of the \emph{excess area}
\begin{equation}\label{excessarea}
\mathcal{A} = \int_{-T}^T dt\, \left[x(t)-x_0(t)\right]
\end{equation}
under the excursion. In the absence of the moving wall, $x_0(t)\equiv 0$, $\mathcal{P}\left(\mathcal{A},T\right)$  coincides with the \emph{Airy distribution}. The Airy distribution exhibits the scaling behavior $\mathcal{P}\left(\mathcal{A}, T\right)=T^{-3/2}\,f(\mathcal{A}/T^{3/2})$ , and the function $f(z)$ is known analytically \cite{Darling,Louchard,Takacs}. The Airy distribution has surprisingly many applications. Most of them belong to computer science \cite{Flajolet,MC}, but the Airy distribution also describes the stationary statistics of the height of a whole class of fluctuating interfaces in one dimension \cite{MC}.

In the presence of a moving wall the probability distribution $\mathcal{P}\left(\mathcal{A},T\right)$ is unknown. Here we calculate the \emph{large deviation function} (LDF) of this  distribution  at $T\to \infty$. As we will see,
the LDF of the excess area has quite interesting properties. To calculate the LDF, we employ
the optimal fluctuation method (OFM), also known as weak noise theory, or WKB theory \cite{FW}.  For the Brownian motion the OFM is essentially the geometrical optics approximation. Using the OFM, we
approximate the probability of observing an atypically large value of the excess area $\mathcal{A}$ by the probability of the \emph{optimal} (that is, most probable) path, or ray $x\left(t\right)$, which escapes the wall and is constrained by Eq.~(\ref{excessarea}). Mathematically, this approximation involves a saddle-point evaluation of the path integral of the properly constrained Brownian excursion. As we argue here, the OFM is asymptotically exact for long times and/or sufficiently large excess areas.

As we will see shortly (see also Ref. \cite{SmithMeerson2018}), the OFM problem of determining the optimal path can be reduced to a simple geometric construction. The resulting optimal path $x\left(t\right)$ is in general composed of parabolic segments and segments of the wall. The OFM uncovers, in a remarkably simple way,  a generic singularity of the LDF, which can be interpreted as a dynamical phase transition of third order. For some wall functions additional phase transitions, which result from a coexistence of multiple OFM solutions,  should be possible. As we argue, these transitions can be quite unusual.
Finally, we will also use the OFM to probe the scaling of \emph{typical} excess area fluctuations with time $T$.

In Sec. \ref{ConstrainedBrownianExcursion} we introduce the model and the OFM. Our calculations are presented in Sec. \ref{main}, and the results are summarized
in Sec. \ref{disc}.

\section{Geometrical optics of constrained Brownian excursion}

\label{ConstrainedBrownianExcursion}

\begin{figure}
\includegraphics[width=0.4\textwidth,clip=]{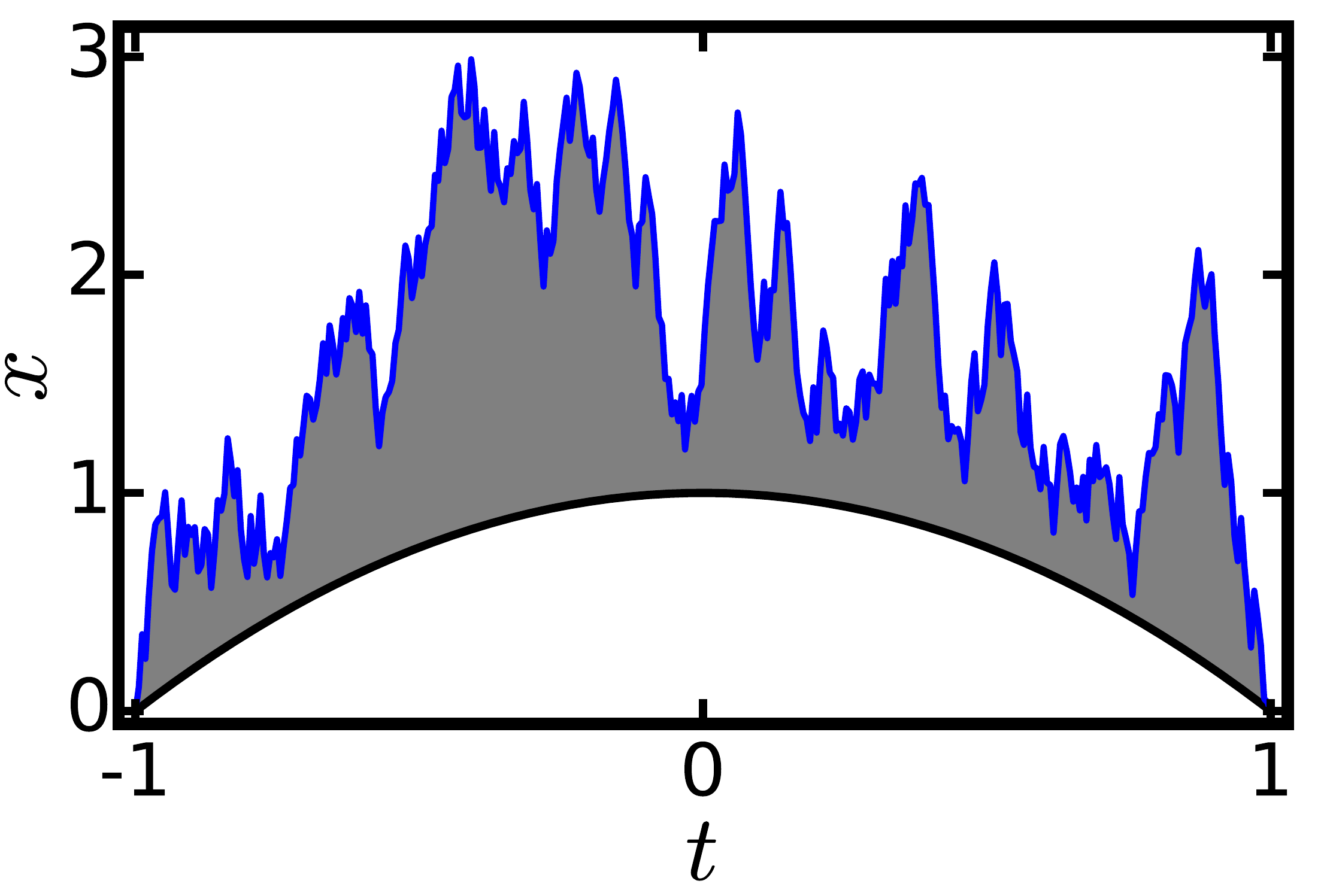}
\caption{A Brownian excursion,  which stays away from the wall moving according to $g\left(t\right) = 1-t^2$. The shaded region has area $\mathcal{A}$, and we are interested in its  distribution. Rescaled units~(\ref{rescaling}) are used.}
\label{realization_example}
\end{figure}

The Brownian motion $x=x\left(t\right)$ can be described by the Langevin equation
\begin{equation}
\label{eq:langevin}
\frac{dx}{dt}=\xi\left(t\right),
\end{equation}
where $\xi$ is a delta-correlated Gaussian noise with zero mean:
\begin{equation}\label{whitenoise}
\left\langle \xi\left(t_{1}\right)\xi\left(t_{2}\right)\right\rangle =2D\delta\left(t_{1}-t_{2}\right).
\end{equation}
The Brownian excursion starts from the point $x=0$ at $t=-T$ and returns
to $x=0$ for the first time at $t=T$. We condition the excursion on staying away from a wall
moving according to the equation
\begin{equation}
x_0(t) = C T^{\gamma} g(t/T) ,
\end{equation}
such that $g\left(\pm1\right)=0$, $g\left(0\right)=1$ and $\gamma > 0$. $C$ is a constant with dimensions length/time$^{\gamma}$.
One realization of this process for the particular case $g(t) = 1-t^2$ is shown in Fig.~\ref{realization_example}. The conditioned trajectories exhibit different excess areas, and we will determine the LDF of the excess area distribution  $\mathcal{P}(\mathcal{A},T)$, where $0<\mathcal{A}<\infty$.

Up to a pre-exponential factor, the unconstrained path probability of the Brownian excursion can be represented as $\propto\text{exp}\left(-S\right)$, where \cite{legacy}
\begin{equation}\label{Action}
S=\frac{1}{4D}\int_{-T}^T \dot{x}^2\,dt.
\end{equation}
The conditional probability distribution $\mathcal{P}(\mathcal{A},T)$ is equal to the ratio of the probabilities of a
wall-escaping Brownian excursion with and without the additional constraint (\ref{excessarea}). Each of these two probabilities can be represented as a path integral over all possible paths [with and without the additional constraint (\ref{excessarea})]. We will assume that each of these path integrals is dominated by the action along a single ``optimal" (or most probable) path, or ray, $x(t)$, for which the action $S$ from Eq.~(\ref{Action}) reaches its minimum. This observation leads to important consequences. To see them, let
us rescale the coordinate $x$ and time $t$:
\begin{equation}\label{rescaling}
 \frac{t}{T} \to t\quad \text{and}\quad \frac{x}{CT^{\gamma}} \to x .
\end{equation}
Upon the rescaling, the condition~(\ref{excessarea}) becomes
\begin{equation}\label{excessarea1}
\int_{-1}^1 dt\, \left[x(t)-x_0(t)\right] = \frac{\mathcal{A}}{C T^{\gamma+1}}\equiv \mathfrak{a},
\end{equation}
whereas the rescaled wall function is simply $g(t)$.  Making the change of variables (\ref{rescaling})
in Eq.~(\ref{Action}) and using Eq.~(\ref{excessarea1}), we uncover the scaling behavior of
the probability density, as predicted by the OFM:
\begin{equation}\label{action1}
-\ln\mathcal{P}(\mathcal{A},T)\simeq \frac{C^2T^{2\gamma-1}}{D}\, s\left(\frac{\mathcal{A}}{CT^{\gamma+1}}\right) .
\end{equation}
It is natural to call the function $s\left(\mathfrak{a}\right)$ the LDF of the distribution $\mathcal{P}(\mathcal{A},T)$. It is given by $s=s_{\text{c}}-s_{\text{u}}$ where $s_{\text{c}}$ and $s_{\text{u}}$ are the rescaled actions,
\begin{equation}
\label{action2}
\frac{1}{4}\int_{-1}^1 \dot{x}^2\,dt\,,
\end{equation}
over the constrained and unconstrained \emph{optimal} paths $x_{\text{c}}\left(t\right)$ and $x_{\text{u}}\left(t\right)$, respectively. Here the ``constrained'' and ``unconstrained'' refer only to the rescaled area constraint (\ref{excessarea1}).

Generally, the OFM is expected to be accurate when it predicts a large action. Equation~(\ref{action1})  implies  that for $\gamma>1/2$ the OFM is asymptotically exact at $T \! \to \! \infty$ provided that the rescaled action $s$ is not too small. For a given $T$, this boils down to a sufficiently large $\mathcal{A}$.

\section{Optimal path and action}

\label{main}

The optimal path of the Brownian excursion must minimize the rescaled action~(\ref{action2}) under the constraint (\ref{excessarea1}). The constraint can be accounted for via a Lagrange multiplier $\lambda$, leading to the effective Lagrangian
$L(x,\dot{x}) = \dot{x}^2/4 - \lambda x$.  The optimal path  $x(t)$
must satisfy the boundary conditions $x(-1)=x(1)=0$ and
stay away from the wall $g(t)$. This leads to a textbook problem of the calculus of variations which deals with one-sided variations \cite{Elsgolts}. The solution typically involves alternating segments of two different types: (1) where
$x (t)$ satisfies the Euler-Lagrange equation $\ddot{x} +2 \lambda = 0$ [so that $x(t)$ is a parabola] and
(2) where $x (t)=g (t)$.  At points where two segments meet they must have a common tangent \cite{Elsgolts}. (The last demand comes from the minimization of the action with respect to the position of the meeting point.) Finally, if there are multiple solutions, the one with the least action must be chosen.

\subsection{Parabolic wall}
\label{sec:parabola}

We will assume throughout this work that the wall function $g(t)$ is smooth and convex upward, $g^{\prime\prime}(t)< 0$, for almost all $|t|\leq 1$. We will also assume for simplicity that $g(-t)=g(t)$. The parabolic wall $g(t)=1-t^2$ is the simplest. Here, for any $\mathfrak{a}>0$, the optimal path -- also a parabola -- is $x(t)=\lambda(1-t^2)$, where $\lambda>1$. The optimal path stays above the wall for all times $|t|\leq 1$.  The rescaled excess area is
\begin{equation}\label{excessparabola}
\mathfrak{a}=\int_{-1}^1 (\lambda-1)(1-t^2)\,dt=\frac{4}{3}\left(\lambda-1\right),
\end{equation}
whereas the rescaled action is
\begin{equation}\label{actionparabola}
s=\frac{1}{4} \int_{-1}^{1} [(-2\lambda t)^2-(-2 t)^2]\,dt = \frac{2}{3} (\lambda^2-1).
\end{equation}
Eliminating $\lambda$ from Eqs.~(\ref{excessparabola}) and (\ref{actionparabola}), we obtain the LDF
\begin{equation}\label{actionparabola1}
s(\mathfrak{a})=\mathfrak{a}+ \frac{3 \mathfrak{a}^2}{8}.
\end{equation}
The resulting probability distribution (\ref{action1}), in the original variables, is
\begin{equation}\label{distparabola}
-\ln\mathcal{P}(\mathcal{A},T)\simeq \frac{C\mathcal{A}}{DT^{2-\gamma}}+\frac{3\mathcal{A}^2}{8DT^3} .
\end{equation}
As one can see, $\mathcal{P}(\mathcal{A},T)$ has two distinct tail asymptotics: the near tail is exponential in $\mathcal{A}$, whereas the far tail is Gaussian. The near tail exhibits the scaling $\mathcal{A} \sim T^{2-\gamma}$.  The presence of the moving wall does not violate the Airy distribution scaling $\mathcal{A}\sim T^{3/2}$ \cite{Darling,Louchard,Flajolet,MC} only in the special case $\gamma=1/2$. We will be mostly interested in $\gamma>1/2$, where the OFM predicts a large action and therefore is accurate.   By analogy with Ref. \cite{SmithMeerson2018} we argue that the scaling  $\mathcal{A} \sim T^{2-\gamma}$ also holds for typical, small fluctuations of $\mathcal{A}$, where the OFM is inapplicable. The reason is quite simple: the near tail identifies uniquely the dimensionless combination of $\mathcal{A}$, $T$, $D$ and $C$ that serves as the dimensionless argument of the probability distribution of typical fluctuations $\mathcal{P}$. The scaling of typical fluctuations of $\mathcal{A}$ with time, up to a numerical coefficient $O(1)$, follows immediately.

The far tail, described by the second term on the right-hand-side of Eq.~(\ref{distparabola}), is wall-independent. It coincides  with the large-$\mathcal{A}$ tail of the Airy distribution (see \textit{e.g.} Ref.  \cite{MC}). This is to be expected: for very large $\mathcal{A}$, when the second term dominates the first one, $\mathcal{P}(\mathcal{A},T)$  is unaffected by the wall. For this reason the Gaussian far tail is universal for all wall functions $g(t)$.

\subsection{Generalized parabolic wall}
\begin{figure}[hb]
\includegraphics[width=0.4\textwidth,clip=]{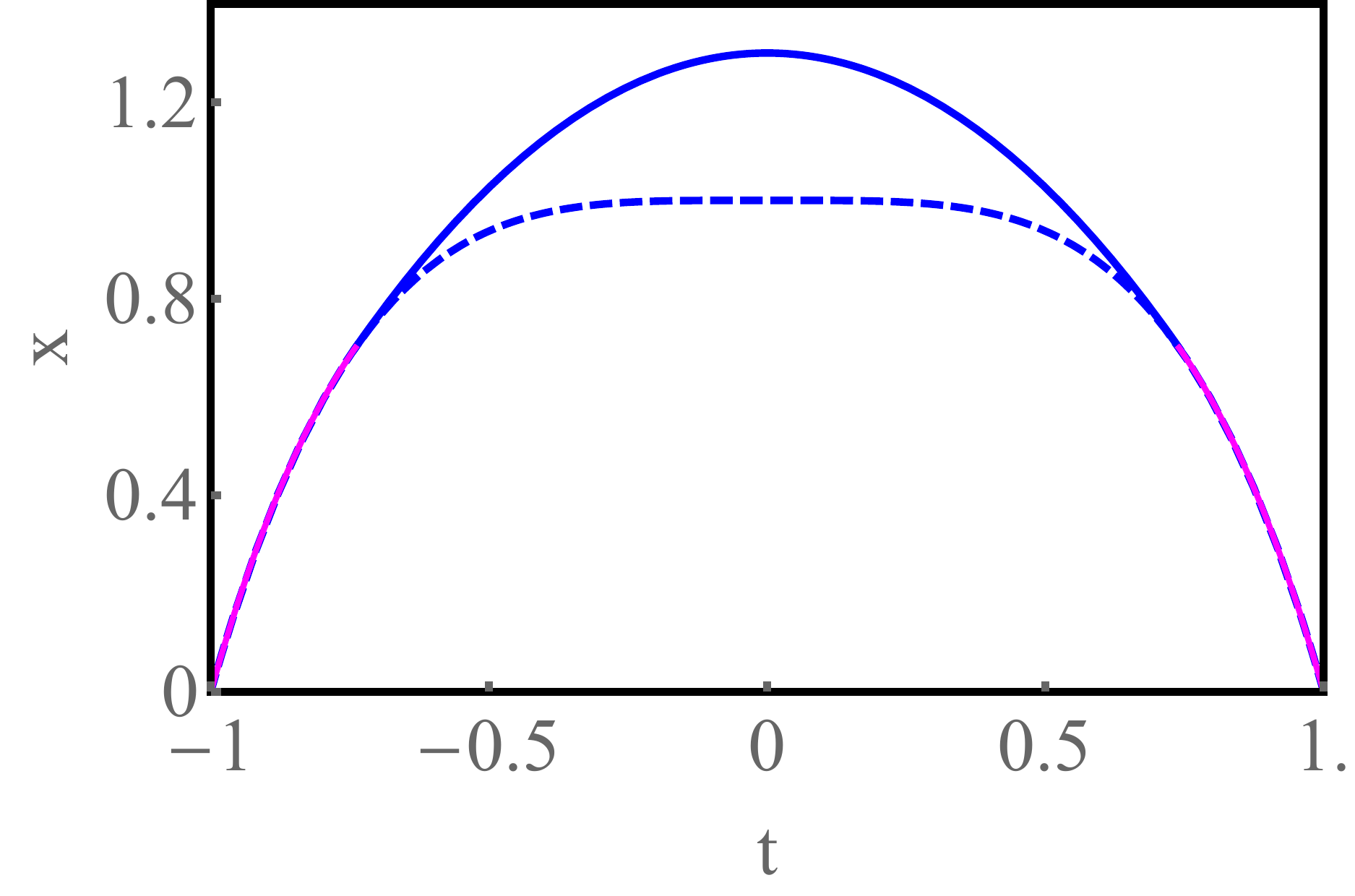}
\includegraphics[width=0.385\textwidth,clip=]{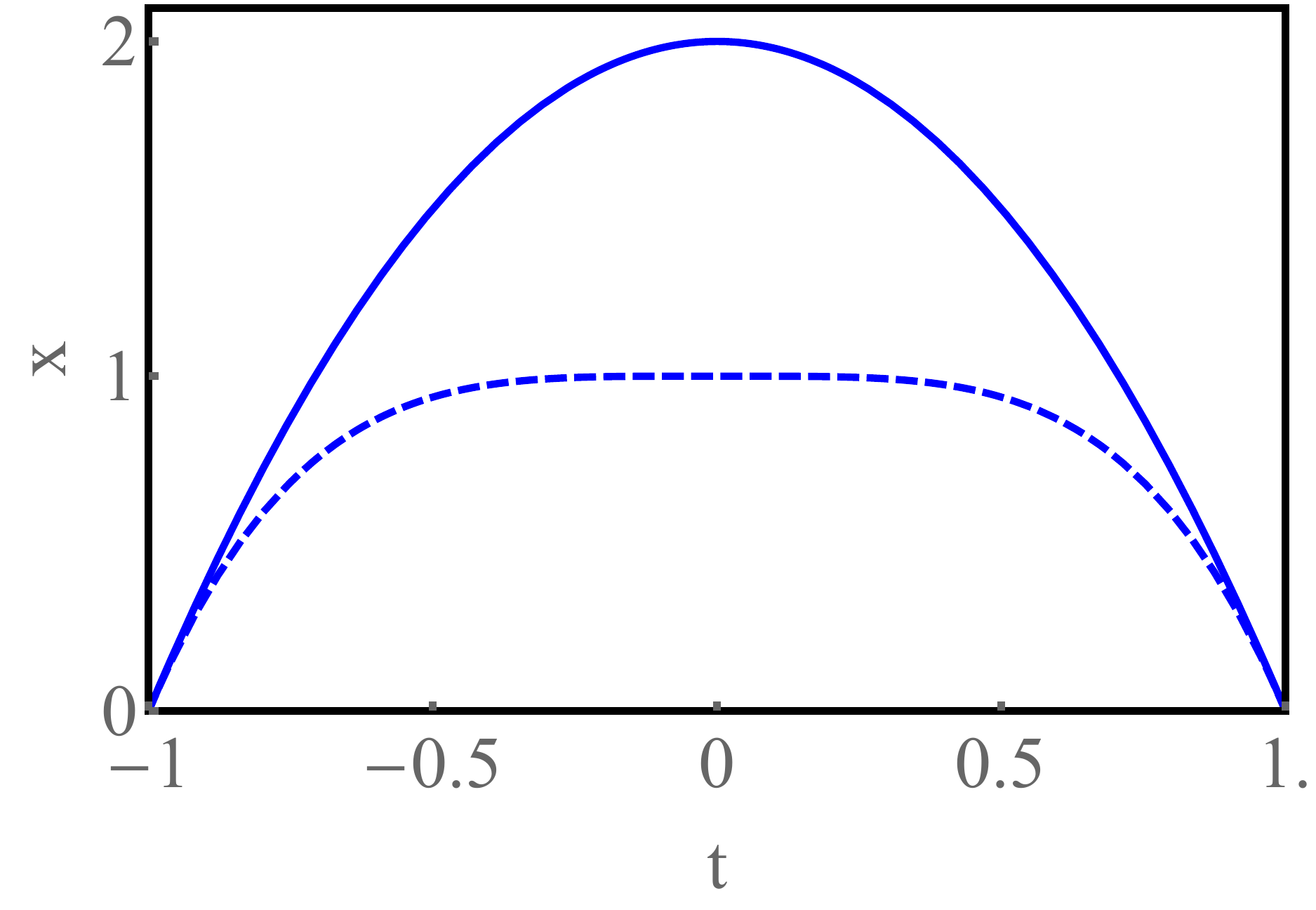}
\caption{The optimal path of Brownian excursion, conditioned on a specified excess area, for the quartic parabola wall function $g(t)=1-t^4$ (shown by the dashed line) below the phase transition (the left panel) and at the transition (the right panel).  Below the transition the optimal path consists of a parabolic segment $x(t)=\lambda(1-t^2)$ (blue solid line) and two wall segments (magenta solid lines). At the transition the optimal path is the parabola $x(t)=2 (1-t^2)$ for all $|t|<1$. Above the transition (not shown) the optimal path is a parabola $x(t)=\lambda (1-t^2)$, with $\lambda>2$, for all $|t|<1$.}
\label{quarticpath}
\end{figure}

A more interesting example is a generalized parabolic wall $g(t)=1-t^{2k}$, where $k>1$ is an integer.  Here the parabolic path $x(t)=\lambda(1-t^2)$ is the optimal path for all $|t|\leq 1$ only when $\lambda>k$, see Fig. \ref{quarticpath}. In this regime of very large deviations of $\mathcal{A}$ the excess area is equal to
\begin{equation}\label{excessgenparabola}
\mathfrak{a}=\int_{-1}^{1}[\lambda(1-t^2)-(1-t^{2k})]\,dt = \frac{4\lambda}{3}-\frac{4k}{2k+1},
\end{equation}
whereas the action is
\begin{equation}\label{actiongenparabola}
s=\frac{1}{4} \int_{-1}^{1} [(-2\lambda t)^2-(-2 k t^{2k-1})^2]\,dt =\frac{2\lambda^2}{3}-\frac{2k^2}{4k-1}.
\end{equation}
Eliminating $\lambda$, we obtain
\begin{equation}\label{actiongenparabola1}
s(\mathfrak{a})\equiv s_{\text{far}}(\mathfrak{a}) =\frac{3 k \mathfrak{a}}{2 k+1}-\frac{8 k^2 (k-1)^2}{(2k+1)^2 (4 k-1)}+\frac{3 \mathfrak{a}^2}{8}.
\end{equation}
As expected, the wall-independent universal term $3\mathfrak{a}^2/8$ dominates at very large $\mathfrak{a}$. Equation~(\ref{actiongenparabola1}) is valid at $\lambda\geq k$, that is at
$$
\mathfrak{a}\geq \mathfrak{a}_{\text{cr}}=\frac{8k(k-1)}{6k+3}.
$$
Now let us consider smaller excess areas, $0<\mathfrak{a}\leq\mathfrak{a}_{\text{cr}}$,
which correspond to $1<\lambda\leq k$. Here the optimal path involves an \textit{a priori} unknown parabolic segment $x(t)=\lambda-bt^2$ which should be matched, at \textit{a priori} unknown times $t=\pm \tau$, with two wall segments by a common tangent construction, see the left panel of Fig. \ref{quarticpath}. After a simple algebra, we obtain
\begin{equation}\label{bandtaugenparabola}
b=k \left(\frac{\lambda
   -1}{k-1}\right)^{\frac{k-1}{k}},\quad \tau=\left(\frac{\lambda
   -1}{k-1}\right)^{\frac{1}{2k}}.
\end{equation}
The excess area is now equal to
\begin{equation}\label{excessgenparabolasmall}
\mathfrak{a}=\int_{-\tau}^{\tau}[\lambda-b t^2-(1-t^{2k})]\,dt = \frac{8 k (\lambda -1)
   \left(\frac{\lambda
   -1}{k-1}\right)^{\frac{1}{2k}}}{6 k+3},
\end{equation}
and the action is
\begin{equation}\label{actiongenparabolasmall}
s=\frac{1}{4} \int_{-\tau}^{\tau} [\dot{x}(t)^2-\dot{g}(t)^2]\,dt = \frac{8 k^2
   (\lambda -1)^{\frac{4k-1}{2
   k}}}{(12 k-3) (k-1)^{\frac{2k-1}{2 k}}}.
\end{equation}
Eliminating $\lambda$, we obtain
\begin{equation}\label{actiongenparabolasmall1}
s(\mathfrak{a})\equiv s_{\text{near}}(\mathfrak{a})=\frac{\left(\frac{3}{8}\right)^{\frac{2k-2}{2k+1}} k^2 \left(\frac{2k+1}{k}\right)^{\frac{4k-1}{2k+1}}}{(4k-1) \left(k-1\right)^{\frac{2k-2}{2k+1}}}\,\mathfrak{a}^{\frac{4k-1}{2k+1}} ,\quad 0<\mathfrak{a}\leq \mathfrak{a}_{\text{cr}}=\frac{8k(k-1)}{6k+3}
\end{equation}
This expression describes a stretched-exponential ``near tail" of $\mathcal{P}(A,T)$.
For $k=1$ (the ``generic" case) the exponential tail $s\sim \mathfrak{a}$ is restored.  Going back to Eq.~(\ref{action1}), we see that the scaling of $\mathcal{A}$ with $T$ in the near-tail region
depends continuously on $k$ and $\gamma$:
\begin{equation}\label{scalingk}
\mathcal{A} \sim   T^{\frac{6k-3\gamma}{4k-1}}, \quad k\geq 1.
\end{equation}
This scaling [which corresponds to a stretched-exponential tail of $\mathcal{P}(\mathcal{A})$] describes  typical fluctuations of $\mathcal{A}$. These fluctuations are determined by the local behavior of the wall function $g(t)$ near $t=0$. This is because, at small $\mathfrak{a}$, $\lambda$ is also small, and  the small optimal parabolic segment $\lambda(1-t^2)$  ``feels" the function $g(t)$  only in a small vicinity at $t=0$.

\begin{figure}[hb]
\includegraphics[width=0.4\textwidth,clip=]{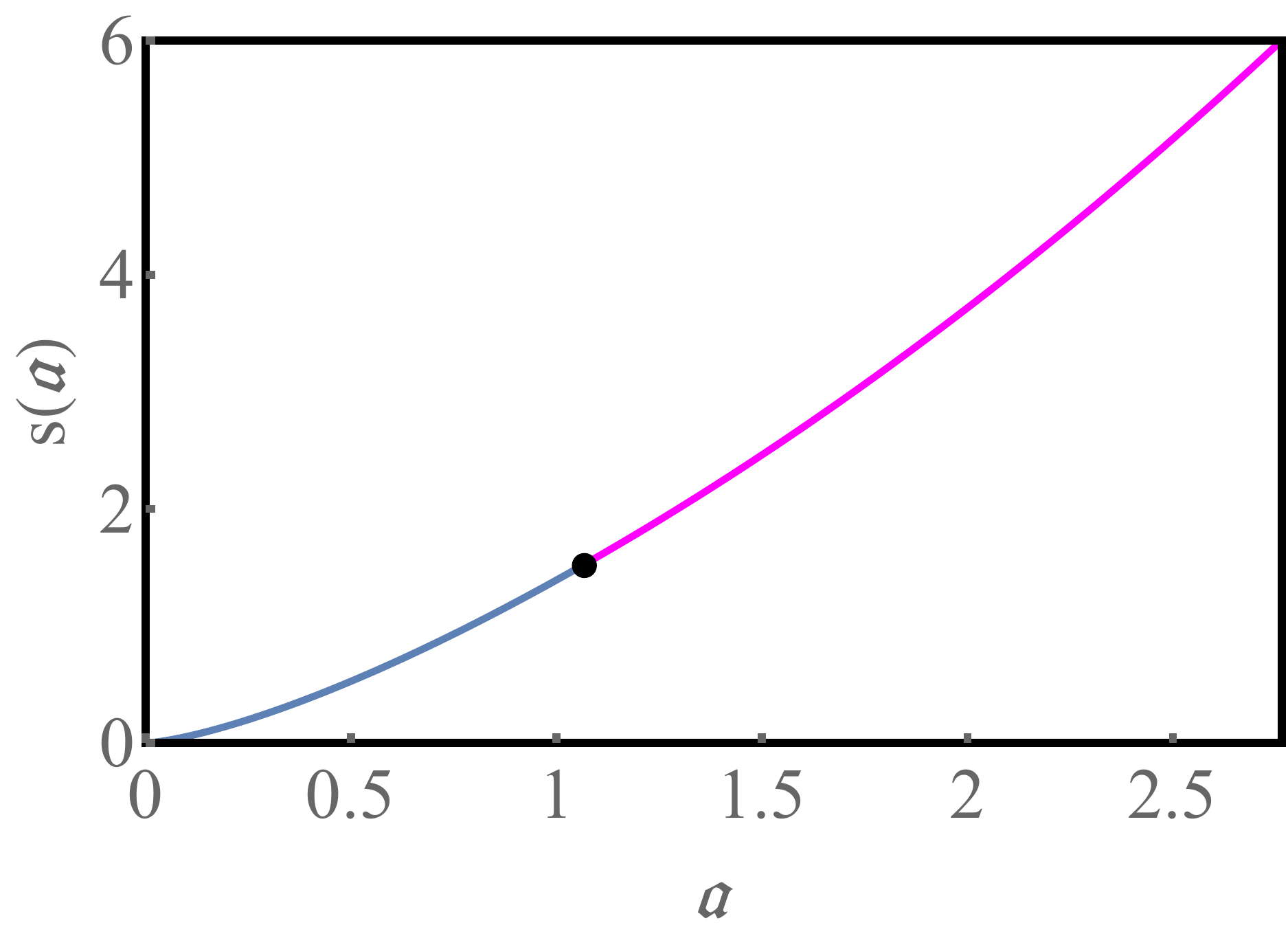}
\caption{The large deviation function of the excess area, $s(\mathfrak{a})$ from Eq.~(\ref{ldfk}), for the quartic parabola wall function $g(t)=1-t^4$. The third derivative $s^{\prime\prime\prime}(\mathfrak{a})$ has a discontinuity at the transition point $\mathfrak{a}=8k(k-1)/(6k+3)=16/15$, shown by the fat circle.}
\label{ldfquarticparabola}
\end{figure}

Overall, the LDF of the excess area is
\begin{equation}
\label{ldfk}
s(\mathfrak{a})\!=\!\begin{cases}
s_{\text{near}}(\mathfrak{a})\;\text{from}\;\text{Eq.}~(\ref{actiongenparabolasmall1}), & 0<\mathfrak{a}\leq\mathfrak{a}_{\text{cr}}=\frac{8k(k-1)}{6k+3},
\\
s_{\text{far}}(\mathfrak{a}) \;\text{from}\;\text{Eq.}~(\ref{actiongenparabola1}), & \mathfrak{a}\geq \mathfrak{a}_{\text{cr}}=\frac{8k(k-1)}{6k+3}.
\end{cases}
\end{equation}
We checked that Eq.~(\ref{ldfk}) actually holds for any $k>1$, not necessarily integer. Figure~\ref{ldfquarticparabola} shows $s(\mathfrak{a})$ for a quartic parabola wall function, $k=2$. At $\mathfrak{a}=\mathfrak{a}_{\text{cr}}$, $s(\mathfrak{a})$ and its first and second derivatives $s^{\prime}(\mathfrak{a})$ and $s^{\prime\prime}(\mathfrak{a})$ are continuous, but the third derivative  $s^{\prime\prime\prime}(\mathfrak{a})$ is discontinuous. This singularity can be interpreted as a third-order dynamical phase transition. The transition occurs when $\tau$ from Eq.~(\ref{bandtaugenparabola}) is equal to $1$, and the order of the transition is determined by the behavior of the wall function $g(t)$ in the vicinity of the end points $t=\pm 1$. For example, for $t$ close to $-1$ one has
$$
g(t)=2 k (t+1)-k (2 k-1) (t+1)^2 +\frac{2}{3} k (k-1) (2 k-1) (t+1)^3+\dots,
$$
As one can see, the coefficient of the $(t+1)^2$ term is nonzero (and negative, so as $g(t)$ is convex upward) for all $k>1$.  As a result, the order of the transition is independent of $k$.

A phase transition of the same type  occurs, at some critical value of $\mathfrak{a}>0$, for many other wall functions. As we have seen, the parabolic wall, $g(t)=1-t^2$, is an exception:  here there is no transition.
The transition is also absent if $g'(t)$ diverges at $t=\pm 1$, as it happens for the circular wall $g(t)=(1-t^2)^{1/2}$.

\subsection{Cosine wall}
\label{sec:cosine}

\begin{figure}[hb]
\includegraphics[width=0.4\textwidth,clip=]{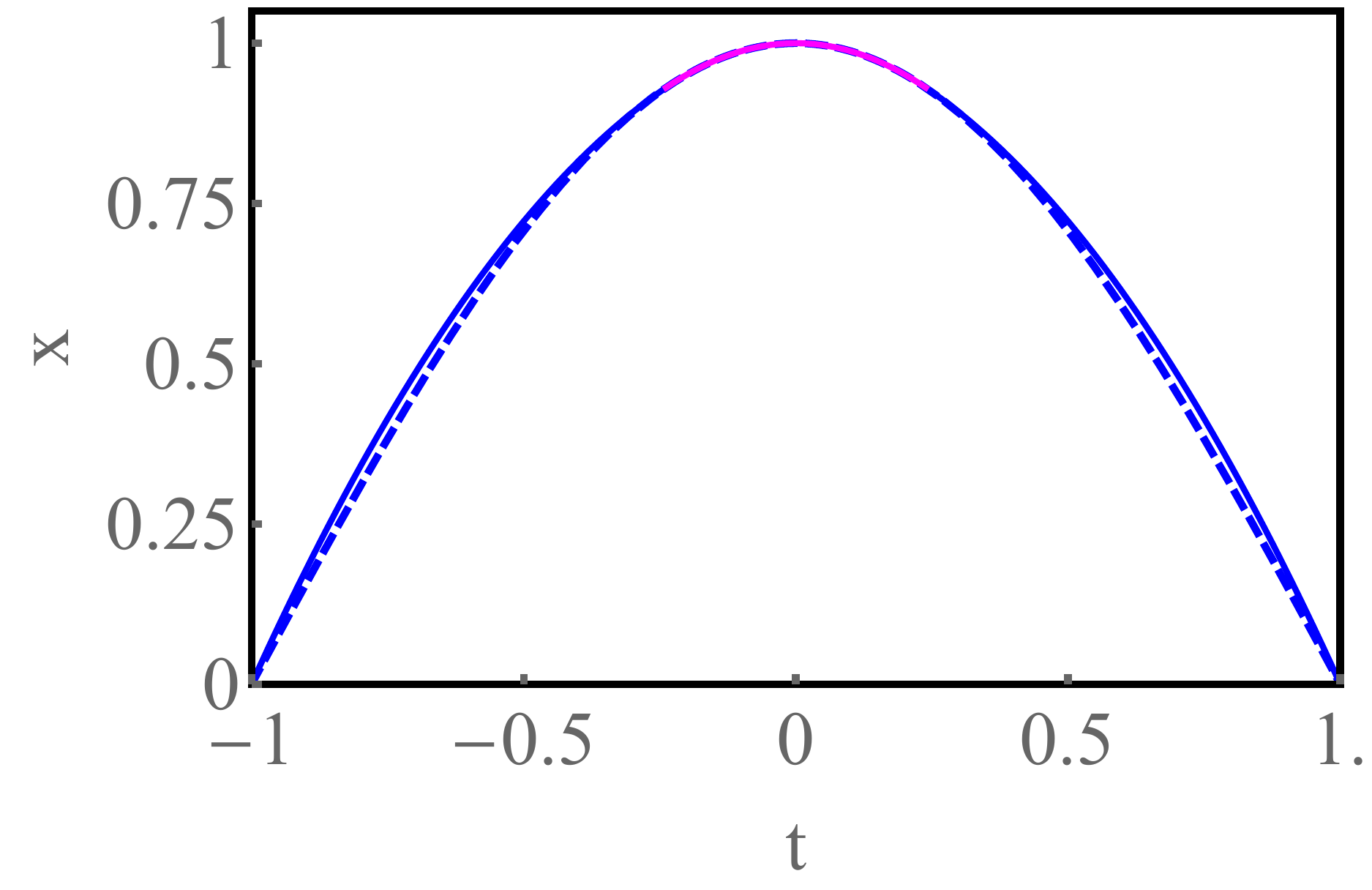}
\includegraphics[width=0.4\textwidth,clip=]{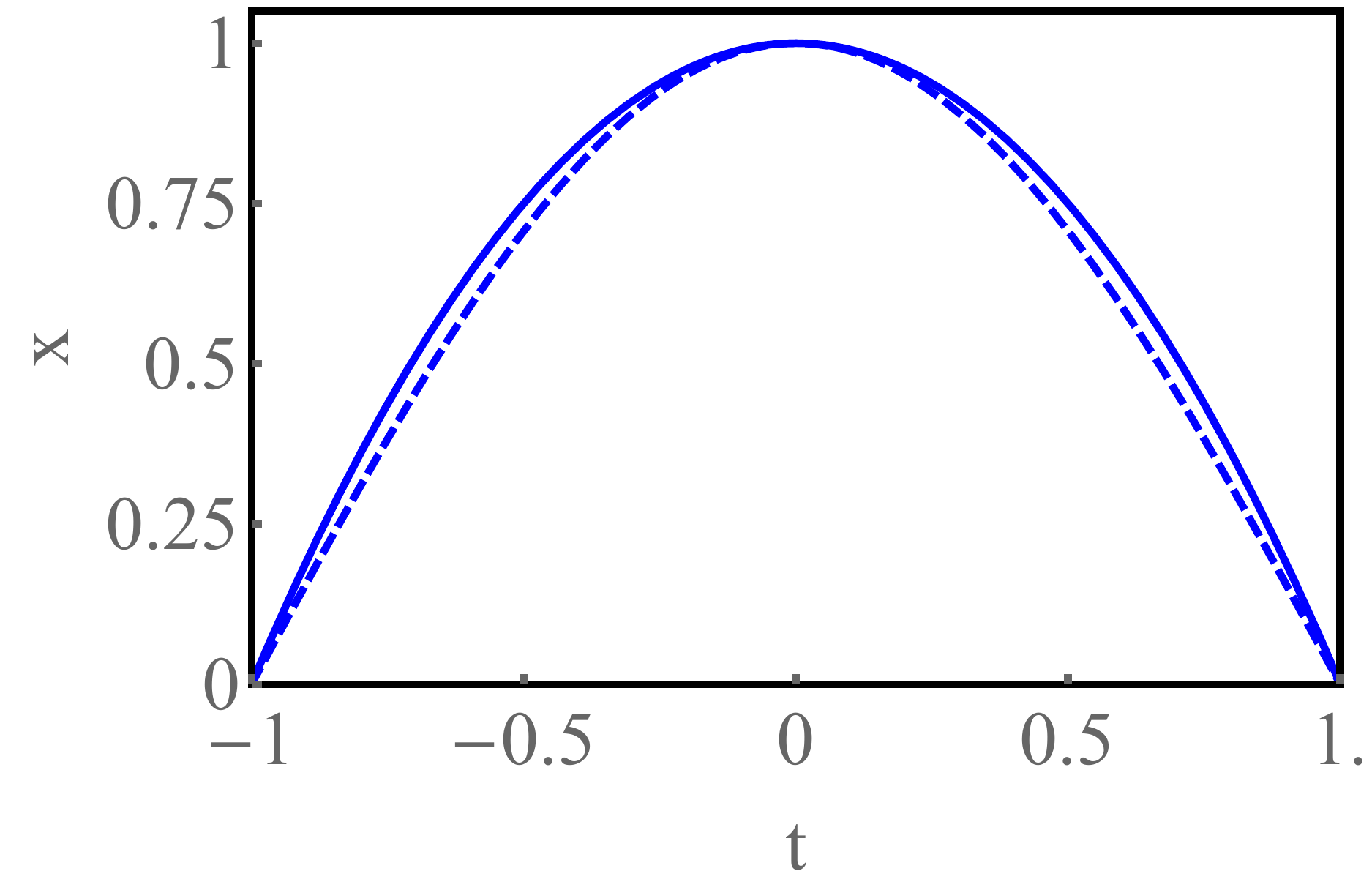}
\caption{The optimal path of Brownian excursion, conditioned on a specified excess area, for the  wall function $g(t)=\cos(\pi t/2)$ (shown by the dashed line) below the phase transition (the left panel) and at the transition (the right panel).  Below the transition the optimal path consists of two parabolic segments (blue solid lines) and one wall segment (magenta solid line). At the transition the optimal path is the parabola $x(t)=1-t^2$ for all $|t|<1$. Above the transition (not shown) the optimal path is a parabola $x(t)=\lambda (1-t^2)$, with $\lambda>1$, for all $|t|<1$.}
\label{cosinepath}
\end{figure}

Above the phase transition, the optimal path is the parabola $x(t)=\lambda (1-t^2)$: for all times $|t|<1$ and for all wall functions. Below the phase transition the situation is more complicated because of different possible mutual arrangements of the optimal parabolic segment(s) of $x(t)$ and the wall function $g(t)$. To illustrate these differences, let us consider the cosine wall $g(t)=\cos(\pi t/2)$. Here the phase transition occurs at $\lambda=1$, which corresponds to
\begin{equation}\label{critcos}
\mathfrak{a}=\mathfrak{a}_{\text{cr}}=\int_{-1}^{1}\left[(1-t^2)-\cos\left(\frac{\pi t}{2}\right)\right]\,dt = \frac{4}{3}-\frac{4}{\pi} = 0.0600937 \dots .
\end{equation}
Above the transition the excess area is
\begin{equation}\label{excesscos}
\int_{-1}^{1}\left[\lambda(1-t^2)-\cos\left(\frac{\pi t}{2}\right)\right]\,dt = \frac{4\lambda}{3}-\frac{4}{\pi}=\mathfrak{a}.
\end{equation}
In its turn, the action is
\begin{equation}\label{actioncos}
s=\frac{1}{4} \int_{-1}^{1} \Bigg\{  (-2\lambda t)^2-\left[-\frac{\pi}{2}\,\sin\left(\frac{\pi t}{2}\right)\right]^2\Bigg\}\,dt =\frac{2\lambda^2}{3} - \frac{\pi^2}{16}, \quad\lambda>1,
\end{equation}
and we obtain
\begin{equation}\label{actioncos1}
s(\mathfrak{a}) =\frac{3 \mathfrak{a}}{\pi} -\frac{\pi ^2}{16}+\frac{6}{\pi^2}+\frac{3 \mathfrak{a}^2}{8},\quad \mathfrak{a}\geq \mathfrak{a}_{\text{cr}}.
\end{equation}
What happens at $0<\lambda<1$, or $\mathfrak{a}<\mathfrak{a}_{\text{cr}}$? For the cosine wall, the coefficient of the $t^4$ term of the Taylor expansion at small $t$,
$$
\cos\left(\frac{\pi t}{2}\right) = 1 -\frac{\pi ^2
   t^2}{8}+\frac{\pi ^4 t^4}{384} - \dots ,
$$
is positive. As a result, a kippa-like optimal path, shown on the left panel of Fig. \ref{quarticpath}, is impossible in this case.  The correct optimal path has two shoulders:  mutually symmetric parabolic segments,  passing through $t=-1$ and $t=1$, respectively, and having common tangents with the wall at some $t=\pm \tau$, see Fig. \ref{cosinepath}. Here the tangent construction can be done numerically. Analytical results can be obtained for  (1) very small $\mathfrak{a}$, which corresponds to the near tail of $\mathcal{P}(\mathcal{A},T)$, and (2) slightly below the phase transition,  $\mathfrak{a}_{\text{cr}}-\mathfrak{a}\ll \mathfrak{a}_{\text{cr}}$.

For  very small $\mathfrak{a}$ the common tangent points $t=\pm \tau$ are very close to $t=\pm 1$, and the wall function $g(t)=\cos(\pi t/2)$ can be Taylor expanded around $t=\pm 1$,
\begin{equation}\label{taylorcos1}
g(t) = \frac{\pi}{2}   (1-|t|) -\frac{\pi^3}{48} (1-|t|)^3+\dots ,
\end{equation}
for the purpose of calculating the common tangent points and the optimal parabola. After some algebra we obtain the near tail
\begin{equation}\label{scossmall}
s=\frac{2^{1/4} \pi^{9/4} \mathfrak{a}^{5/4}}{5 \sqrt{3}},\quad \mathfrak{a}\ll \mathfrak{a}_{\text{cr}}.
\end{equation}
This stretched exponential tail is determined by the local properties of the wall function at $t=\pm 1$, rather than at $t=0$. The unusual exponent $5/4$ appears because there is no quadratic term $O[(1-|t|)^2]$ in the Taylor expansion (\ref{taylorcos1}). When a quadratic term is present, one obtains $s \sim \mathfrak{a}$, leading to an exponential near tail.

Slightly below the phase transition,  $\mathfrak{a}_{\text{cr}}-\mathfrak{a}\ll \mathfrak{a}_{\text{cr}}$, the common tangent points $t=\pm \tau$ are very close to zero, and we can Taylor expand the wall function $g(t)=\cos(\pi t/2)$ there:
\begin{equation}\label{taylorcos2}
g(t) =1 - \frac{\pi ^2 t^2}{8}+\frac{\pi ^4 t^4}{384}+\dots .
\end{equation}
In the first order in $\mathfrak{a}_{\text{cr}}-\mathfrak{a}$, the $t^4$ term can be neglected, and the calculations are very simple.  In this order the result is
\begin{equation}\label{scoslarge}
s(\mathfrak{a})\simeq \frac{4}{\pi} -\frac{2}{3}- \frac{\pi^2}{16}  + \mathfrak{a} ,\quad \mathfrak{a}_{\text{cr}}-\mathfrak{a}\ll \mathfrak{a}_{\text{cr}}.
\end{equation}
At $\mathfrak{a}=\mathfrak{a}_{\text{a}}$ this expression matches, together with its first derivative, with the asymptotic (\ref{actioncos1}). To prove that the transition is of third order (as we conjecture), one would need to continue the calculations until the third order in $\mathfrak{a}_{\text{cr}}-\mathfrak{a}$.

\subsection{Multiple solutions and additional phase transitions}
\label{sec:general}

For  a class of wall functions both types of the parabolic optimal paths (the kippa-like and the shoulders-like) are possible for the same value of $\mathfrak{a}$. This situation can occur when the Taylor expansions of $g(t)$ at $t=0$ and $t=\pm 1$ have the following forms:
\begin{equation}
\label{taylorgen}
g(t)\!=\!\begin{cases}
1-b t^2-c t^4 +\dots, & \quad |t|\ll 1,
\\
A (1-|t|)-B (1-|t|)^2-C (1-|t|)^3+\dots, & \quad 1-|t|\ll 1.
\end{cases}
\end{equation}
and all the coefficients $b,c, A, B$ and $C$ are positive\footnote{The coefficients $A$, $B$ and $b$ are always positive for a $g(t)$ which is convex upward.}.  A direct calculation shows that, for the kippa-like path, the leading-order action at small $\mathfrak{a}$ is  $s_1(\mathfrak{a})\simeq b\,\mathfrak{a}$, whereas for the shoulders-like path it is $s_2(\mathfrak{a}) \simeq B\,\mathfrak{a}$. As the correct solution must minimize the action, the selected optimal path ``nucleates" at $t=0$ (if $b<B$) or at $t=\pm 1$ (if $b>B$).  The corresponding near tail of $\mathcal{P}(\mathcal{A},T)$ is exponential, with $s=\text{min} (b,B) \,\mathfrak{a}$.

\begin{figure}[hb]
\includegraphics[width=0.4\textwidth,clip=]{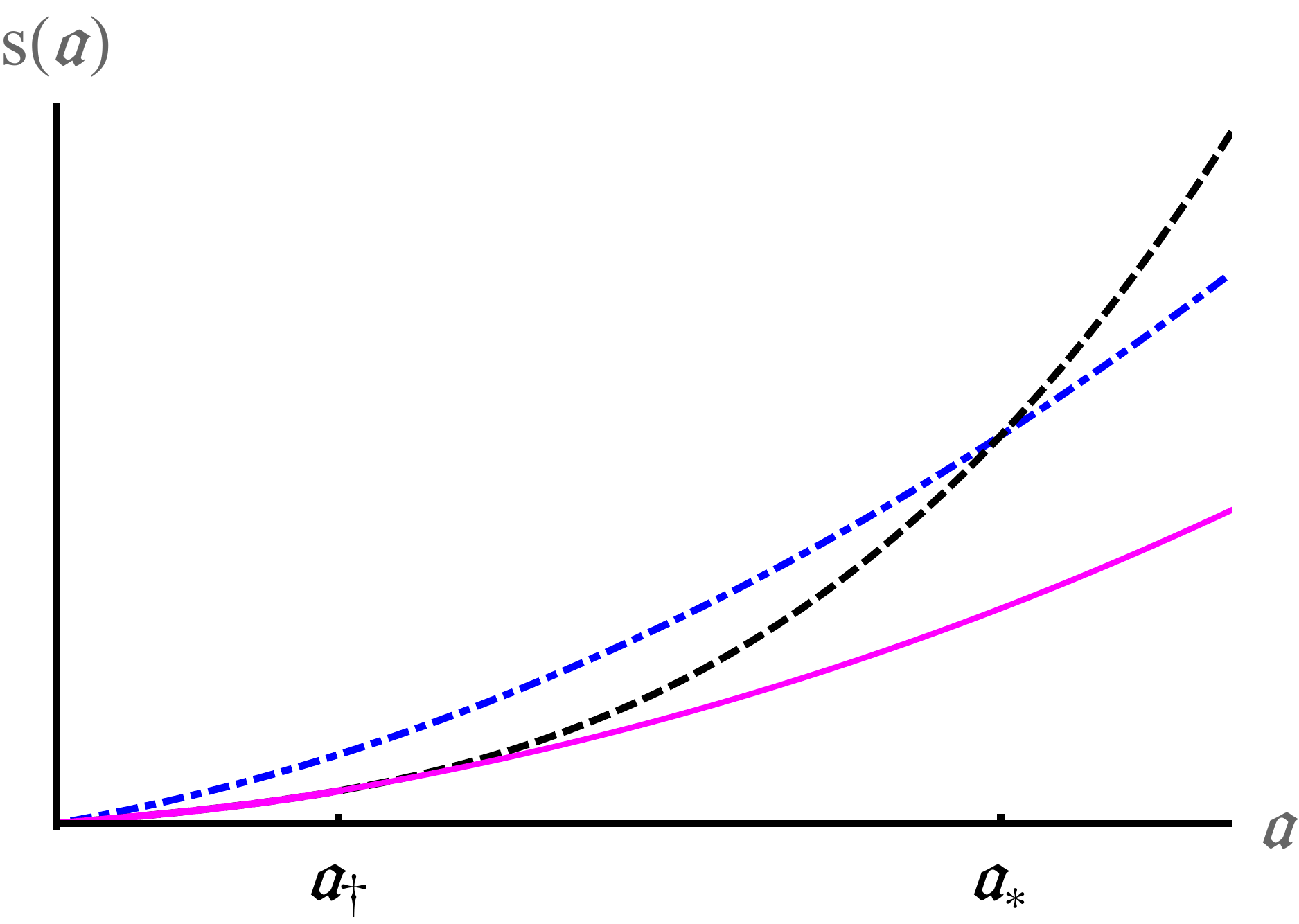}
\caption{A phase transition, originating from an interplay of two coexisting OFM solutions: the kippa-like and the shoulders-like. Shown by the dashed and dash-dotted lines are the functions  $s_1(\mathfrak{a})$ and $s_2(\mathfrak{a})$, respectively. The resulting LDF  $s(\mathfrak{a})$  is shown by the solid line. The critical point $\mathfrak{a}_{\dagger}$ is the common tangent point of $s(\mathfrak{a})$  and $s_1(\mathfrak{a})$.
The critical point $\mathfrak{a}_{\text{cr}}$ (where the kippa-like and the two shoulders-like paths merge) is at larger $\mathfrak{a}$ and not shown.}
\label{firstorderfig}
\end{figure}

As $\mathfrak{a}$ increases, the functions $s_1(\mathfrak{a})$ and $s_2(\mathfrak{a})$ become affected by higher-order terms in the expansions~(\ref{taylorgen}). It can happen  that, for some $\mathfrak{a}_*$ (which is subcritical
with respect to the third-order transition considered above),
one has $s_1(\mathfrak{a})<s_2(\mathfrak{a})$ for $\mathfrak{a}<\mathfrak{a}_*$, but $s_1(\mathfrak{a})>s_2(\mathfrak{a})$ for $\mathfrak{a}>\mathfrak{a}_*$. Na\"{\i}vely, one would expect a jump in the first derivative $s'(\mathfrak{a})$ at  $\mathfrak{a}=\mathfrak{a}_*$, see the dashed and dash-dotted lines in Fig. \ref{firstorderfig}.
The correct LDF $s(\mathfrak{a})$, however, is quite different.
In order to calculate it one should minimize the sum $s_1(\mathfrak{a}_1)+s_2(\mathfrak{a}_2)$ with respect to $\mathfrak{a}_1$ and $\mathfrak{a}_2$ under the constraint $\mathfrak{a}_1+\mathfrak{a}_2=\mathfrak{a}$. The result is schematically shown by the solid line in Fig. \ref{firstorderfig}. Importantly, $s(\mathfrak{a})$ (which is \emph{not} a linear function) has only one common tangent with $s_1(\mathfrak{a})$, at some point $\mathfrak{a}=\mathfrak{a}_{\dagger}<\mathfrak{a}_*$.  At $0<\mathfrak{a}\leq\mathfrak{a}_{\dagger}$ the LDF $s(\mathfrak{a})$ coincides with $s_1(\mathfrak{a})$. At $\mathfrak{a}>\mathfrak{a}_{\dagger}$, however, $s(\mathfrak{a})$ is smaller than any of the functions $s_1(\mathfrak{a}_1)$ and $s_2(\mathfrak{a}_2)$.  At $\mathfrak{a}=\mathfrak{a}_{\dagger}$ $s(\mathfrak{a})$ has a jump in its second derivative with respect to  $\mathfrak{a}$. That the system avoids  a first-order transition at $\mathfrak{a}=\mathfrak{a}_*$ and instead exhibits a \emph{single} second-order transition at $\mathfrak{a}=\mathfrak{a}_{\dagger}<\mathfrak{a}_*$ is a new and unexpected feature.

In fact, this argument can be pushed further.  Figure~\ref{firstorderfig} makes it obvious that, for a second-order transition at $\mathfrak{a}=\mathfrak{a}_{\dagger}$ to occur, the curves $s_1(\mathfrak{a})$ and $s_2(\mathfrak{a})$ do not even need to cross each other at a point $\mathfrak{a}_*<\mathfrak{a}_{\text{cr}}$. In other words, this second-order transition can happen even without an attempted first-order transition.

\section{Summary and discussion}

\label{disc}

The OFM is very efficient in its description of a broad class of atypically large fluctuations. Therefore, it may come as a surprise that the advantages of the OFM have not been sufficiently appreciated in the context of constrained Brownian motions and their applications. We started filling this gap in Ref.  \cite{SmithMeerson2018} and continued doing it in the present work. Here we calculated the large deviation function (LDF) of the excess area $\mathcal{A}$ of a Brownian excursion, constrained to stay away from a moving wall. For a whole class of walls, the LDF has a jump in the third derivative with respect to $\mathcal{A}$ at a critical value of $\mathcal{A}$.
It is natural to interpret this singularity as a dynamical phase transition. The transition mechanism -- a space-time ``obstacle",  experienced by the ``diffusion ray" -- is remarkably simple.

The OFM allows us to probe the scaling behavior $\mathcal{A}\sim T^{\alpha}$ of typical (small) fluctuations of the excess area by evaluating their distribution tail (which we call the near tail). The scaling exponent $\alpha$ depends continuously on the parameters $\gamma$ and $k$ which characterize the moving wall.

One surprising outcome of this work is that, for some wall functions, additional phase transitions are possible, which result from coexistence of different optimal paths, predicted by the OFM. We identified the mechanism of one such transition, of the second order.
It would be very interesting to investigate these phase transitions in more detail. It would be also interesting to calculate (with a different method) the probability distribution of \emph{typical} fluctuations of the excess area $\mathcal{A}$, and see how their distribution match in the tail with the large deviations, considered in this work.

Before we finish, let us return to the third-order transition uncovered in this work. In recent years third-order transitions have been identified in large deviation functions characterizing a whole list of stochastic many-body systems, see  Ref. \cite{shortreview}  for an illuminating review. These include Gaussian random matrices, non-intersecting Brownian excursions in one dimension, nonequilibrium stochastic growth models  belonging to the Kardar-Parisi-Zhang  universality
class \cite{shortreview,3orderKPZ}, \textit{etc}. The common features of these third-order transitions are the following \cite{shortreview}:
\begin{itemize}
\item{The region of typical fluctuations is ``sandwiched" between two large-deviation tails.}
\item{The large-deviation tails scale differently with a large parameter $N\gg 1$ of the problem; and the sharp  transition appears when $N\to \infty$.}
\item{The typical fluctuations are described by the Tracy-Widom distribution \cite{TW}.}
\end{itemize}
The third-order transition, that we uncovered in this work, looks different on all counts:
\begin{itemize}
\item{The region of typical fluctuations is located outside of the transition point.}
\item{The large-deviation tails have identical  scaling behaviors with $T\gg 1$ below and above the transition.}
\item{The typical fluctuations are \emph{not} described by the Tracy-Widom distribution.}
\end{itemize}
On the other hand, this transition has a simple geometric mechanism which is apparently not shared by the transitions described above. This is a good instance to ponder about universality, or a lack of thereof.

\section*{ACKNOWLEDGMENTS}

I am very grateful to Naftali Smith for valuable advice and for producing Fig. 1. I acknowledge a useful discussion with Tal Agranov. I am also grateful to the Center of Mathematical Research (Centro di Ricerca Matematica) Ennio De Giorgi in Pisa, where this work started, for hospitality.
This research was supported by the Israel Science Foundation (grant No. 807/16).

\end{document}